\documentclass[preprint]{aastex}
\RequirePackage{epstopdf}

\bibpunct{(}{)}{;}{a}{}{;}

\begin{document}
\received{ }
\accepted{  }
%\journalid{ }{ }
%\articleid{ }{ }
%\slugcomment{   }
\slugcomment{Accepted by ApJ on 07 August 2012}

\shortauthors{Adams et al.}
\shorttitle{SOFIA/FORCAST Imaging of W3(OH)}

\title{SOFIA/FORCAST and Spitzer/IRAC Imaging of the Ultra Compact H {\small II} Region W3(OH) and Associated Protostars in W3}

\author{Lea Hirsch\altaffilmark{1}, Joseph D. Adams\altaffilmark{1}, Terry L. Herter\altaffilmark{1}, Joseph L. Hora\altaffilmark{2}, James M. De Buizer\altaffilmark{3}, S. Thomas Megeath\altaffilmark{4}, George E. Gull\altaffilmark{1}, Charles P. Henderson\altaffilmark{1}, Luke D. Keller\altaffilmark{5}, Justin Schoenwald\altaffilmark{1}, William Vacca\altaffilmark{3}}

\altaffiltext{1}{Cornell University, Department of Astronomy, 105 Space Sciences Bldg., Ithaca, NY 14853, USA}
\altaffiltext{2}{Harvard-Smithsonian Center for Astrophysics, 60 Garden St., MS 65, Cambridge, MA 02138-1516, USA}
\altaffiltext{3}{SOFIA-Universities Space Research Association, NASA Ames Reseach Center, Mail Stop N211-3, Moffett Field, CA 94035, USA}
\altaffiltext{4}{University of Toledo, Department of Physics and Astronomy, Mailstop 111, 2801 West Bancroft Street, Toledo, Ohio 43606, USA}
\altaffiltext{5}{Ithaca College, Physics Department, 264 Ctr for Natural Sciences, Ithaca, NY 14850, USA}

\begin{abstract}
We present infrared observations of the ultra-compact H {\small II} region W3(OH) made by the FORCAST instrument aboard SOFIA
and by Spitzer/IRAC. We contribute new wavelength data to the spectral energy distribution, which constrains the optical depth,
grain size distribution, and temperature gradient of the dusty shell surrounding the H {\small II} region.
We model the dust component as a spherical shell containing an inner cavity with radius $\sim 600$ AU, irradiated
by a central star of type O9 and temperature $\sim 31,000$ K. The total luminosity of this system is $7.1 \times 10^4~L_\odot$. 
An observed excess of $2.2 - 4.5~\mu$m emission in the SED can be explained by our viewing a cavity opening or clumpiness
in the shell structure whereby radiation from the warm interior of the shell can escape.
We claim to detect the nearby water maser source W3 (H$_2$O) at 31.4 and $37.1~\mu$m using beam deconvolution of the FORCAST images. We constrain the flux densities of this object at $19.7 - 37.1~\mu$m.
Additionally, we present {\it in situ} observations of four young stellar and protostellar objects in the SOFIA field, presumably associated
with the W3 molecular cloud. Results from the model SED fitting tool of \citet{Robitaille06,Robitaille07} suggest that 
two objects (2MASS J02270352+6152357 and 2MASS J02270824+6152281) are intermediate-luminosity 
($\sim 236 - 432~L_\odot$)  protostars;
one object (2MASS J02270887+6152344) is either a high-mass protostar with luminosity $3\times10^3$ L$_{\odot}$ or 
a less massive young star with a substantial circumstellar disk but depleted envelope; 
and one object (2MASS J02270743+6152281) is an intermediate-luminosity ($\sim 768~L_\odot$) protostar
nearing the end of its envelope accretion phase or a young star surrounded by a
circumstellar disk with no appreciable circumstellar envelope.
\end{abstract} 


\section{INTRODUCTION}\label{Introduction}
Ultra-compact H {\small II} regions can be found throughout the galaxy surrounding newly formed massive stars just completing the accretion stage and entering
their main sequence lifetimes. UCHs are small ($<0.4$~pc), hot ($>100$~K), and massive ($>10^{3} M_{\odot}$), and emit $10^{4} - 10^{6}~L_\odot$~
\citep{Churchwell}. These regions of ionized gas are surrounded by molecular and dust clouds, out to as far as ten times the radius of the UCHs themselves
\citep{Conti}. This causes attenuation of much of the emitted luminosity, as it is absorbed and reradiated in the infrared.

W3(OH) is one of the largest and best studied UCH {\small II} regions known in the Galaxy. 
%The work of \citet{Dreher} indicates that the 
%UCH {\small II} region in W3(OH) may be as young as 270 years, meaning its O-type exciting star is only just reaching its main sequence %lifetime. 
It is at a distance of approximately 2.04 kpc in the Perseus arm \citep{Hachisuka}. 
Like most UCH \small{II} regions, is surrounded by its natal dust and molecular gas envelope as well as a larger giant molecular cloud encompassing 
W3(OH) and W3(H$_{2}$O), W3 Main, and AFGL 333; these are all regions of potentially triggered star formation, 
based on their positions on the dense outskirts 
of a much less dense cavity in the W3 GMC \citep{Ruch, Moore}. Further review of star formation in the W3 GMC is presented by \citet{Megeath}.

Line emission from the molecular gas surrounding W3(OH) has been mapped at submillimeter and radio wavelengths in the transitions of 
OH and H$_2$O \citep{mader78},  HCN \citep{Turner}, NH$_3$  \citep{wilson78, zeng84, tieftrunk98}, CH$_3$OH \citep{Menten}, 
C$^{18}$O \citep{Wink}, and C$^{17}$O \citep{Wyrowski97}. This molecular emission is found out to a $1'$ diameter region. 
These studies have particularly focused on further characterizing the molecular gas and the OH and methanol masers within and around 
W3(OH), as well as the hot core/water maser region W3(H$_{2}$O) approximately 6'' to the east.  In continuum emission, however, the dust cocoon 
surrounding the UCH {\small II} region comes into primary focus. 

The dust component to the UCH {\small II} W3(OH) was detected by \citet{wynnwilliams72} at $\sim 1 - 20~\mu$m; the
dust component is optically thick in the near- and mid-infrared. \citet{chini86} studied the cold dust at wavelengths of
350 $\mu$m and 1.3 mm.
Using a spherically symmetric radiative transfer model, they concluded that the inner cavity ($\lesssim 2 \times 10^{17}$~cm) 
of the UCH {\small II} region is depleted of dust, rather than having dust density that increases approaching the central star. 
A large amount of visual extinction (A$_{\rm v} \sim 67$) in a thick outer shell was required to explain the decline in emission at 
$\lambda \lesssim 5~\mu$m. Utilizing airborne data, \citet{campbell89} showed
that the dust cocoon is optically thick in the far-infrared and developed a more detailed model containing an H {\small II} region and a cavity.
Surrounding the cavity is a dusty region with a free-fall density distribution and a temperature gradient through the dust cloud. More recently, \citet{Stecklum} presented high 
spatial resolution,
ground-based 10 and 20 $\mu$m images of W3(OH). From these data, the model for the dust shell was further developed and contained
a Gaussian density distribution with an inner cavity of radius 2270 AU and stellar luminosity of $8 \times 10^4~L_\odot$. 

In this work, we present new, high spatial resolution observations of the W3(OH) region in the wavelength range $\sim 3.6 - 40~\mu$m obtained
with the FORCAST instrument \citep{herter12} on SOFIA \citep{young12} and with the Infrared Array Camera (IRAC; Fazio et al. \citeyear{fazio04})
on the Spitzer Space Telescope \citep{werner04}.
These wavelengths are critical for determining the spectral energy distribution (SED) of the W3(OH) dust cocoon and thus for making
a measurement of its total luminosity. We combine our data with 2MASS \citep{Skrutskie06} fluxes 
and other data published in the literature to construct SEDs for the W3(OH) dust component. 
We model the dust component as a dusty shell around the H {\small II} region and compute the emergent flux
using radiative transfer code. 
In addition, we present {\it in situ} observations of four stellar and protostellar objects in the SOFIA/FORCAST field.
We fit the SEDs of these objects to those of high- and intermediate-mass protostars containing dusty circumstellar envelopes
and circumstellar disks. We discuss the physical properties of all these objects.

\section{OBSERVATIONS}\label{Observations}
\subsection{SOFIA/FORCAST Observations}
Observations were conducted on December 8, 2010 aboard the Stratospheric Observatory for Infrared Astronomy (SOFIA), using the Cornell-built Faint Object Infrared
Camera for the SOFIA Telescope (FORCAST) \citep{Adams}. FORCAST uses a Si:As BIB detector array at wavelengths $\lambda \le 26 \mu$m and a Si:Sb BIB array at
$\lambda \ge 26 \mu$m. The detector has a $0.7638''$ pixel scale (rectified) over a total field of view of $3.4' \times 3.2'$. Four filters were utilized for these
observations, with central wavelengths (and bandpasses) at 19.7 (5.5), 24.2 (2.9), 31.4 (5.7), and 37.1 (3.3) $\mu$m. The dichroic beamsplitter
was used to obtain the wavelength pairs 19.7/37.1 and 24.2/31.4 $\mu$m. Asymmetrical chopping with respect to the optical
axis was applied in order to avoid telescope coma in the on-source beam.
The chop throw was $5'$ and was designed to chop off nearby nebulosity.
A 5-point dither pattern in C2NC2 mode \citep{herter12} was implemented to remove bad pixels during post-processing. 
Integration times were 30 sec at every dither position which yielded a total of 150 sec in each filter configuration.
The data were pipeline processed with the reduction algorithms described in \citet{herter12}. 
A global solution to the calibration from the 
SOFIA Early Science phase was applied to the images \citep{herter12}. We discuss flux extraction of sources in \S\ref{Results}.
Finally, we applied color corrections
to the extracted fluxes as computed by an instrument model with model atmosphere \citep{herter12}.

%Fluxes for the region at each wavelength were extracted photometrically using a $15''$ aperture radius for W3(OH), 
%[and what for the 
%a $4.5''$ aperture radius for 
%J02270881+6152345 and J02270843+6152294, and a $3''$ aperture for J02270729+6152279. We applied color corrections
%to the fluxes as computed by an instrument model with model atmosphere \citep{herter12}.

\subsection{Spitzer/IRAC Observations}
We used data obtained with IRAC on Spitzer.  The data were
taken from observations that were obtained in high dynamic range (HDR)
mode, whereby two images are taken in succession at 0.4s and 10.4s
integration times. The brightest objects in the field, including W3(OH)
were saturated in the longer frames, so we used the 0.4s exposure time
frames to construct the mosaic that was used for the bright source
photometry. The following AORIDs were used: 5050624, 19305728, 20590592,
38744064, 38757632, 38763776, 38769408, 38770944, 38790656, and 38801408.
We utilized the Basic Calibrated Data (BCD) version S18.5 products from the
Spitzer Science Center (SSC) standard data pipeline. For the 3.6 and 4.5
$\mu$m bands, data from the cryogenic and warm mission were combined.
For the 5.8 and 8.0 $\mu$m BCDs, bright source artifacts such as ``banding''
\citep{Hora} were removed using the IMCLEAN image processing
routines\footnotemark\footnotetext[6]{http://irsa.ipac.caltech.edu/data/SPITZER/docs/dataanalysistools/tools/contributed/irac/imclean/}.
 The images were mosaiced with IRACproc \citep{schuster06} which uses a
version of the mopex mosaicing software \citep{makovoz05} developed at the
SSC.  The final mosaic was made with a pixel scale of $0.863''$/pixel.

\section{RESULTS}\label{Results}

\subsection{Imagery and SOFIA/FORCAST detections}
\begin{figure}
\plotone{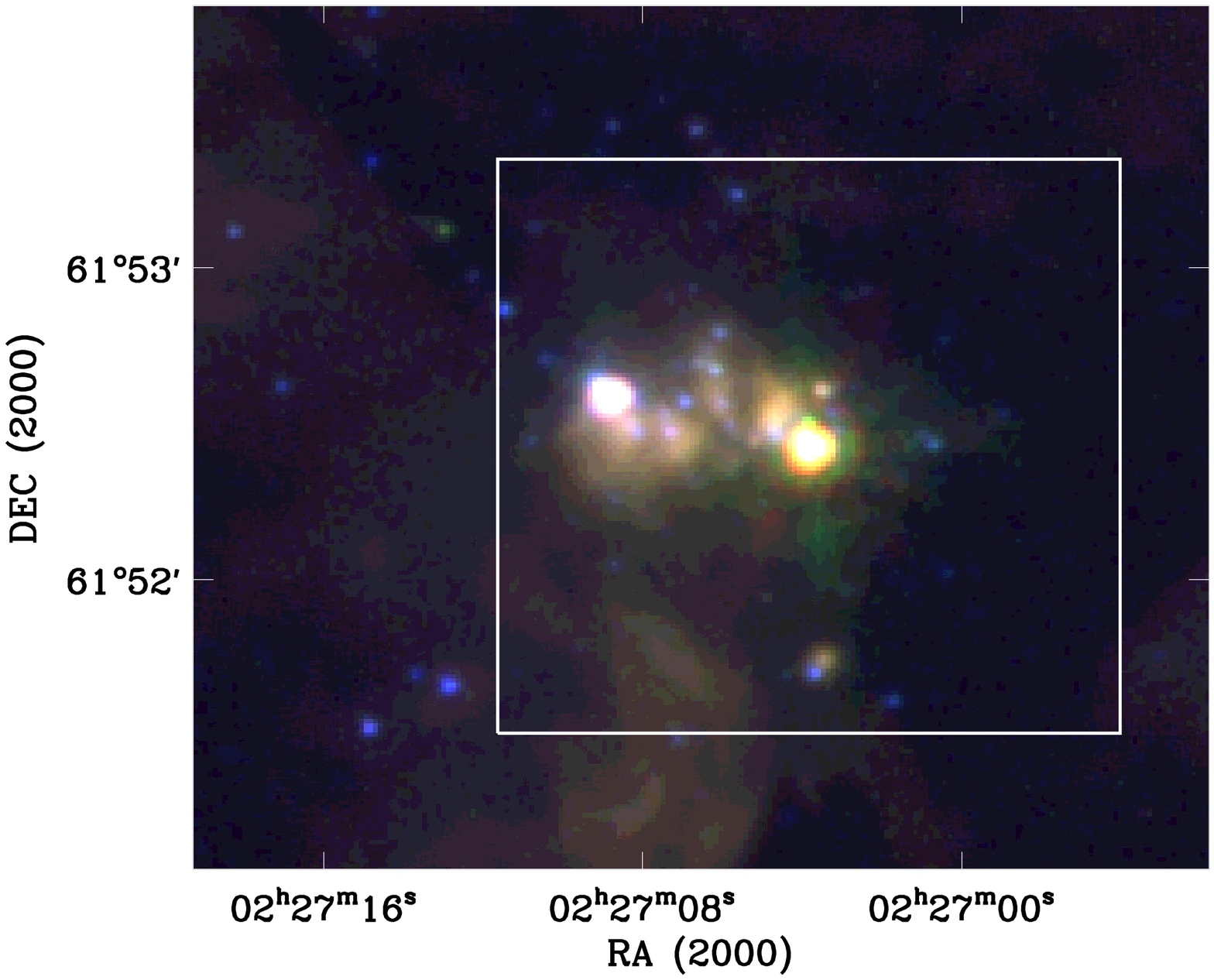}
\caption{False-color IRAC image of the W3(OH) region at 3.6 $\mu$m (blue), 5.8 $\mu$m (green), and 8.0 $\mu$m (red). 
The box indicates the region of the SOFIA field that is shown in Fig. \ref{fig:forcast_image}. 
\label{fig:irac_image}}
\end{figure}

\begin{figure}
\plotone{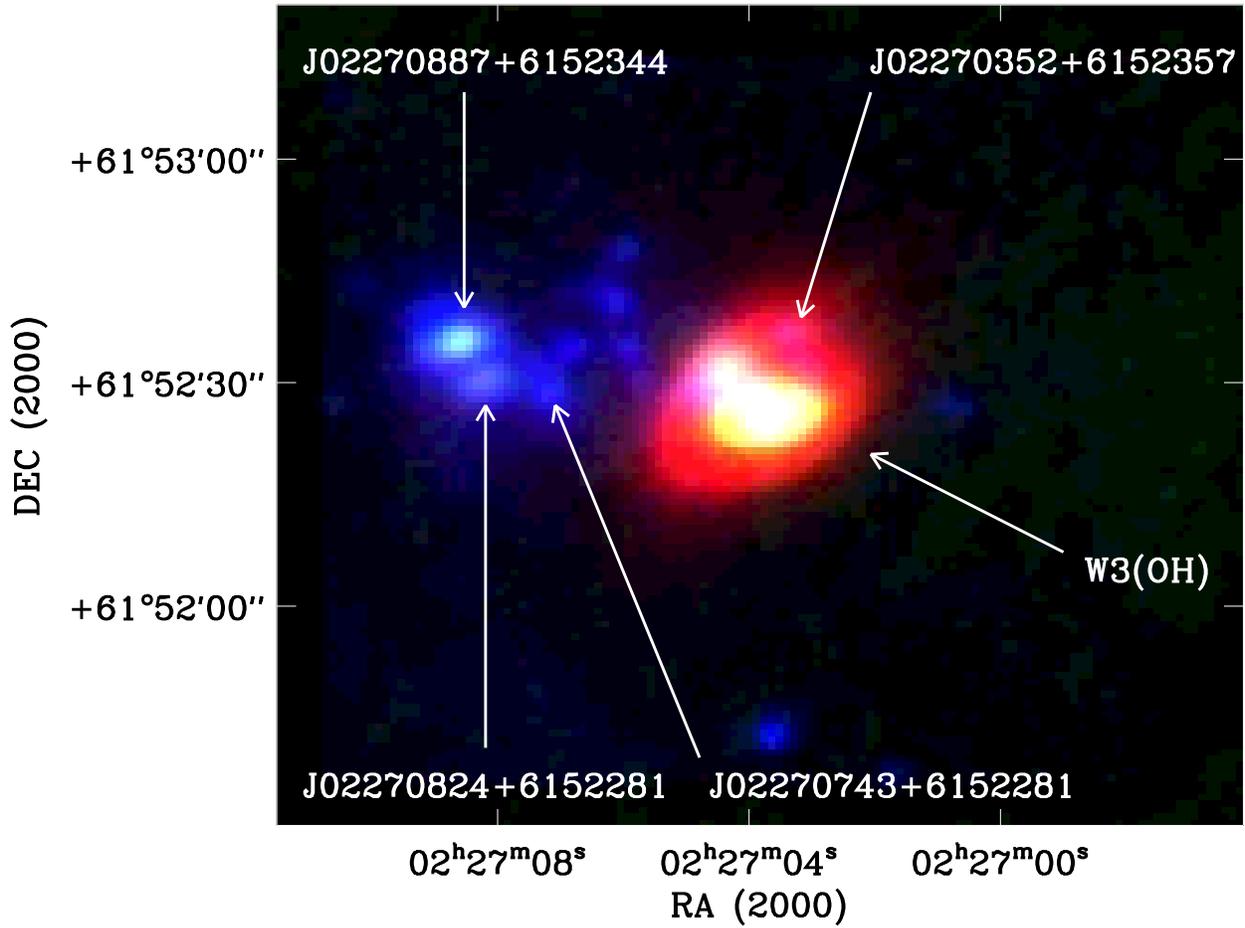}
\caption{False-color image of the W3(OH) region at 3.6 $\mu$m (blue, IRAC), 
19.7 $\mu$m (green, FORCAST), and 37.1 $\mu$m (red, FORCAST). The image at 3.6 $\mu$m was convolved with a
Gaussian kernel to match the spatial resolution of FORCAST at 37.1 $\mu$m. The positions of W3(OH) and the four point
sources that were detected by SOFIA/FORCAST are indicated.
\label{fig:forcast_image}}
\end{figure}

Reduced, multi-wavelength IRAC and FORCAST images are shown in Figs. \ref{fig:irac_image} and \ref{fig:forcast_image}. 
The measured image quality in the FORCAST images was consistent with the overall
image quality achieved by SOFIA during the Early Science period \citep{herter12}. W3(OH) is spatially resolved
in the FORCAST images.  
Within W3(OH), the dual peaks in flux correspond to a bright cometary Southwestern H {\small II} region and a dimmer,
more eliptically-shaped Northeastern H {\small II} region \citep{Stecklum}.

We detect four mid-infrared point sources in the SOFIA field. 
These sources are designated in the 2MASS catalog as J02270352+6152357, J02270743+6152281, J02270824+6152281, and 
J02270887+6152344. There are also $2~\mu$m counterparts
to these sources in the $K'$ band image of \citet{tieftrunk98}. However, there is a set of
rich clusters of low mass stars in this region, so confusion with neighboring
sources precludes our making an accurate measurement of their $2~\mu$m fluxes.
In Fig. \ref{fig:forcast_image}, we label the sources with
their corresponding designations. All four sources were detected by SOFIA/FORCAST at 19.7, 24.2, and 31.5 $\mu$m; 
J02270352+6152357, J02270824+6152281, and J02270887+6152344 were also detected at 37.1 $\mu$m.
In addition, all four sources were detected in the Spitzer
images at wavelengths of 3.6, 4.5, 5.8, and 8.0 $\mu$m; however, the 8.0 $\mu$m detection for J02270743+6152281
suffers from contamination by relatively bright nebulosity extending from below the source.
The fluxes for J02270743+6152281, J02270824+6152281, and 
J02270887+6152344 were extracted from FORCAST images by fitting Gaussian functions to flux line profiles 
and from Spitzer images by aperture photometry. The corresponding flux densities are given in Table \ref{tab:fluxes}. 
Flux extraction for J02270352+6152357 is discussed in \S\ref{sec:deconvolution}.

\subsection{Deconvolved FORCAST images}\label{sec:deconvolution}
We performed beam deconvolution on each of the FORCAST images in order to search 
for the hot core W3(H$_2$O) and to measure the size of W3(OH).
The images were deconvolved using the maximum likelihood method
\citep{richardson72, lucy74}. Like all deconvolution methods, knowledge of
the point-spread function (PSF) of an unresolved source is needed at each
wavelength. The delivered PSF can change due to different wind loads on
the secondary mirror and differences in the telescope flexure as a
function of telescope position. To mitigate these effects on ground-based
telescopes, high S/N observations of mid-infrared bright stars are usually
taken immediately before and/or after each science target observation and
as close to the science target as possible ($<1^\circ$ away) so as to get
the best PSF calibration for use in the deconvolution procedure. However,
in our case, finding a PSF star that is bright enough at wavelengths out
to 40 $\mu$m that fulfills these requirements is nearly impossible and thus was
not attempted. Standard star images taken throughout our flight similarly
did not have sufficient S/N for use as a PSF calibrator at the longer
wavelengths. Therefore, we used these standard stars observations to
determine an average FWHM for each wavelength for the flight. Then
artificially generated PSFs (using an Airy pattern calculated from
the wavelength, telescope diameter, and central obscuration diameter) were
constructed and convolved with a Gaussian to achieve PSFs with FWHMs that
equaled the measured average FWHMs of the standard stars. These idealized
PSFs were then used in the deconvolution procedure. These deconvolved
images compare favorably to simple unsharp masking of the original images,
and hence all of the substructures revealed in the deconvolved images are
believed to be real with high confidence. Final image resolutions are
about a factor of two better than the natural resolutions for each image.

\begin{figure}
\plotone{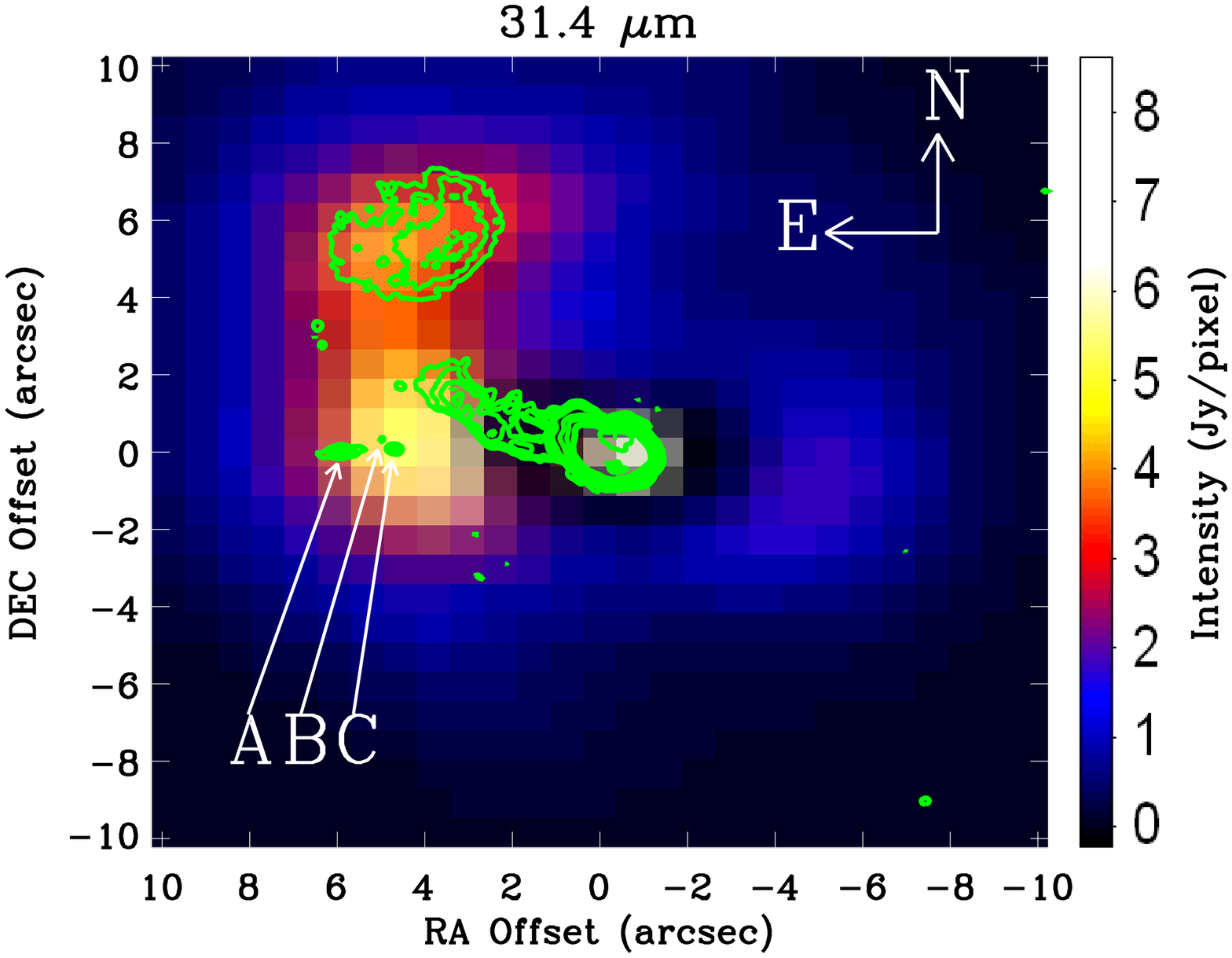}
\caption{Deconvolved SOFIA/FORCAST image of W3(OH) at 31.4 $\mu$m, scaled
linearly in flux. The image is centered on
the coordinates RA=$02^{\rm{h}}27^{\rm{m}}03^{\rm{s}}.87$ and DEC=$+61^\circ52'24''.6$ (J2000).
The core of W3(OH) has been subtracted using a 2-dimensional Gaussian fit to its radial brightness
profile.
The locations of sources A, B, and C in W3(H$_2$O) from \citet{Wyrowski99} are shown.
The green contours represent 8.4 GHz VLA observations from \citet{Wilner}. The contours levels are
0.00004, 0.0001, 0.00018, 0.0003, 0.0005, 0.001, 0.005, 0.01, 0.015, 0.016, 0.02, 0.022, and 0.03 Jy/beam. 
An assumption is that the peak of the 8.4 GHz contours coincides with the peak of the $19.7 - 37.1~\mu$m 
emission. Also seen is the H {\small II} region to the Northeast of the peak of W3(OH).
\label{fig:deconvolved_images_31}}
\end{figure}

\begin{figure}
\plotone{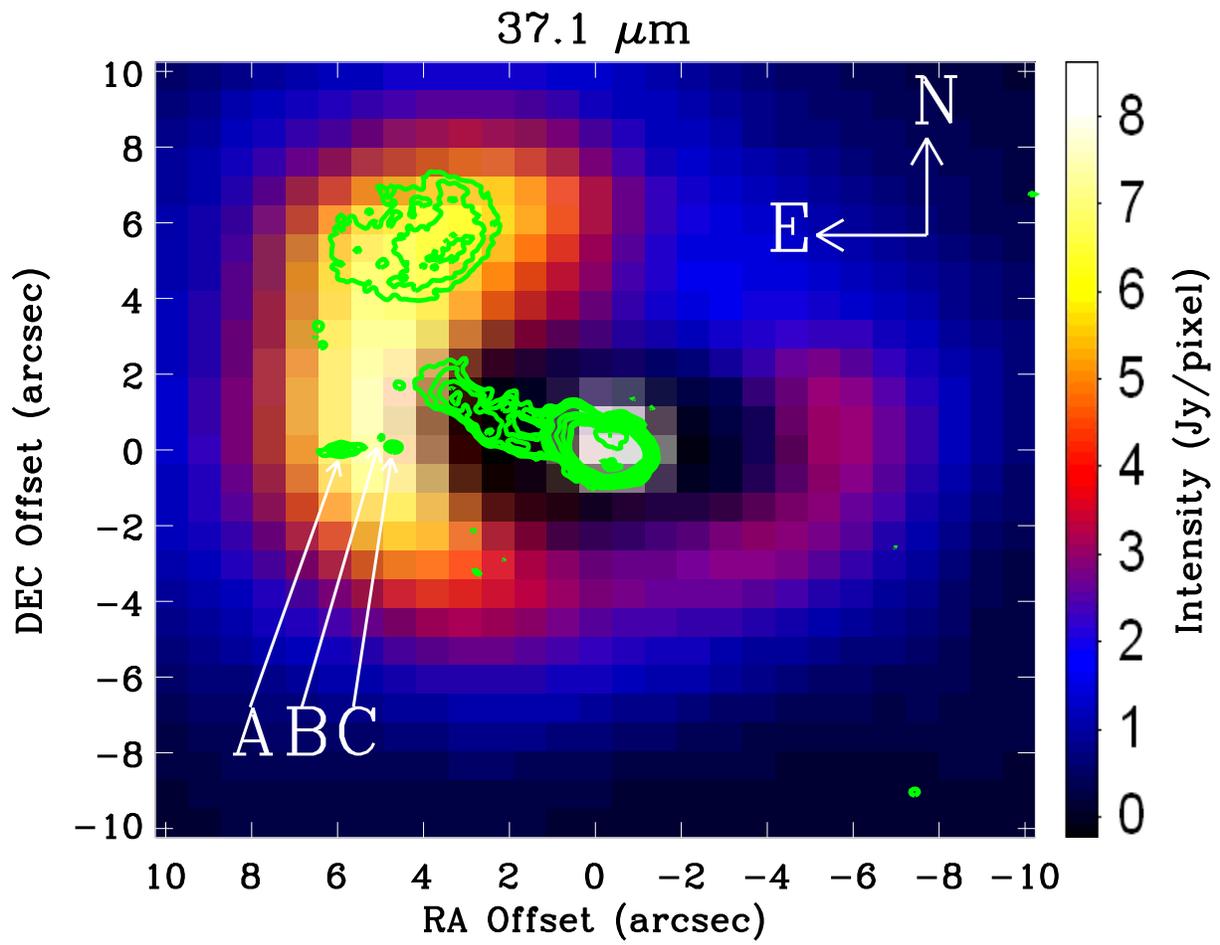}
\caption{Same as Fig. \ref{fig:deconvolved_images_31} for 37.1 $\mu$m.
\label{fig:deconvolved_images_37}}
\end{figure}

We claim to detect unresolved emission from the hot core W3(H$_2$O) at 31.4 and $37.1~\mu$m.
Figs. \ref{fig:deconvolved_images_31} and \ref{fig:deconvolved_images_37} show the region
near W3(H$_2$O) in the deconvolved images at 31.4 and 37.1 $\mu$m, respectively. In each case,
the central core of W3(OH) has been fit to a 2-dimensional Gaussian function and subtracted
from the respective image. 
Figs. \ref{fig:deconvolved_images_31} and \ref{fig:deconvolved_images_37} also contain 
8.4 GHz VLA observations from \citep{Wilner}. The radio
continuum image shows a cometary component extending in the Northeast direction from the peak of W3(OH). The radio
image also shows emission from the Northeast H {\small II} region \citep{Stecklum} above the cometary component. 
In the 8.4 GHz continuum, the W3(H$_2$O) clumps A, B, and C \citep{Wyrowski99} are resolved. 
Figs. \ref{fig:deconvolved_images_31} and \ref{fig:deconvolved_images_37} show residual emission
at 31.4 and 37.1 $\mu$m in the vicinity of W3(H$_2$O). Moreover, the position of this residual emission shifts in wavelength from
close to W3(H$_2$O) C at 31.4 $\mu$m towards W3(H$_2$O) A at 37.1 $\mu$m.

\begin{figure}
\plotone{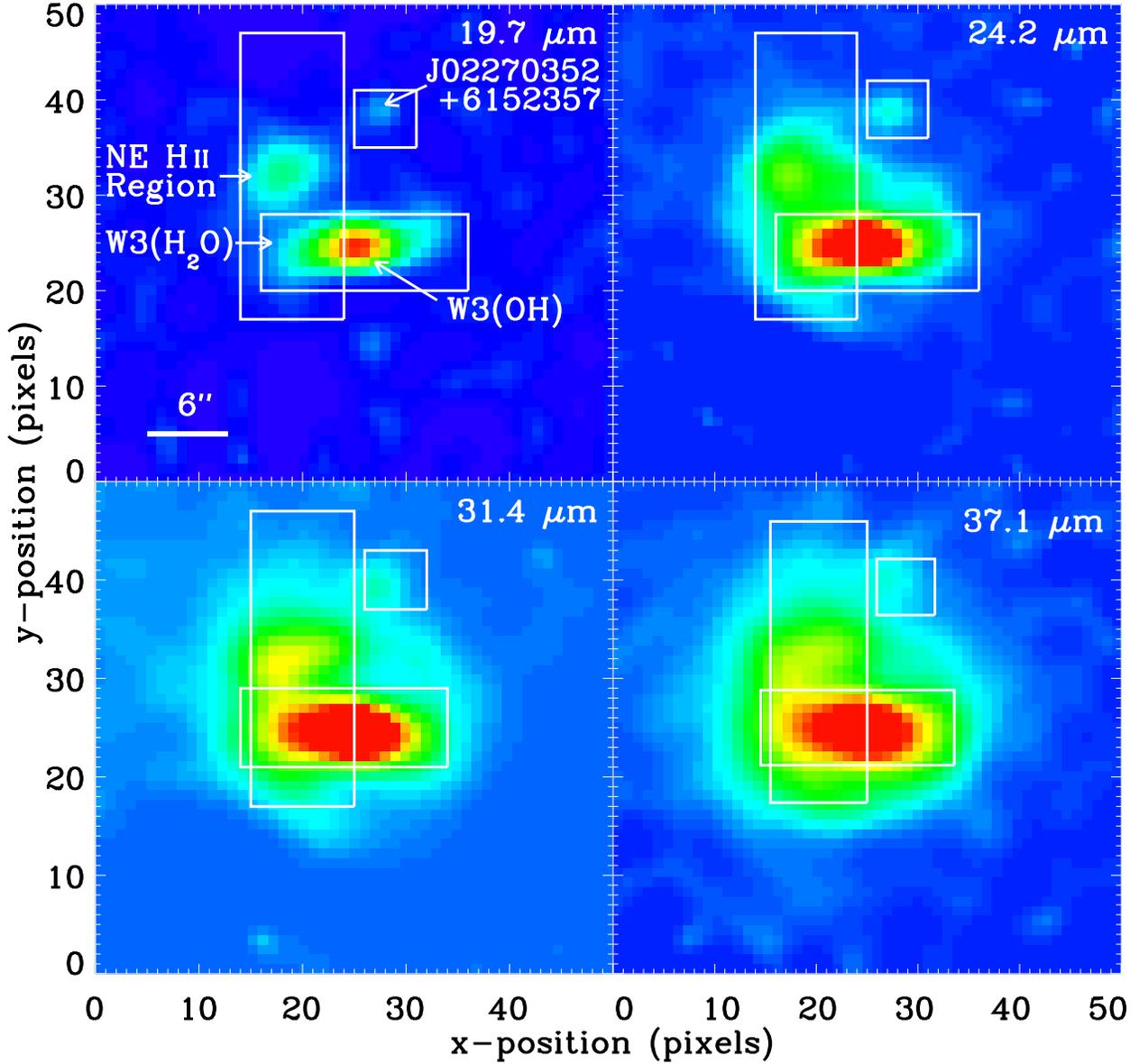}
\caption{SOFIA/FORCAST deconvolved images at 19.7, 24.2, 31.4, and 37.1 $\mu$m, scaled logarithmically in flux with
rainbow color intensity mapping.
The locations of W3(OH), W3(H$_2$O), the Northeastern H {\small II} region \citep{Stecklum}, and
the point source 2MASS J02270352+6152357 are indicated in the upper left panel. The boxed regions
were the regions selected for flux extraction; 1-dimensional integrations
were performed over the short dimensions of the boxes and the resulting line profiles are shown in Fig. \ref{fig:line_plots}.
\label{fig:deconvolved_images_extract}}
\end{figure}

Flux extraction for W3(OH), W3(H$_2$O), the Northeast H {\small II} region, and 2MASS J02270352+6152357 was
performed using the deconvolved images.
The resolved diameter of W3(OH) in the deconvolved images (Fig. \ref{fig:deconvolved_images_extract})
is approximately 40 pixels, corresponding to $\sim 63000$ AU. 
We used a 15-pixel aperture radius to extract the flux density at 19.7 $\mu$m and 20-pixel aperture radii to extract the flux densities at 
24.2, 31.4, and 37.1 $\mu$m; we then subtracted the flux densities
measured for the Northeastern H {\small II} region and 2MASS J02270352+6152357 (discussed next).
The resulting integrated flux densities for W3(OH) are listed in Table \ref{tab:fluxes}.

\begin{figure}
\plotone{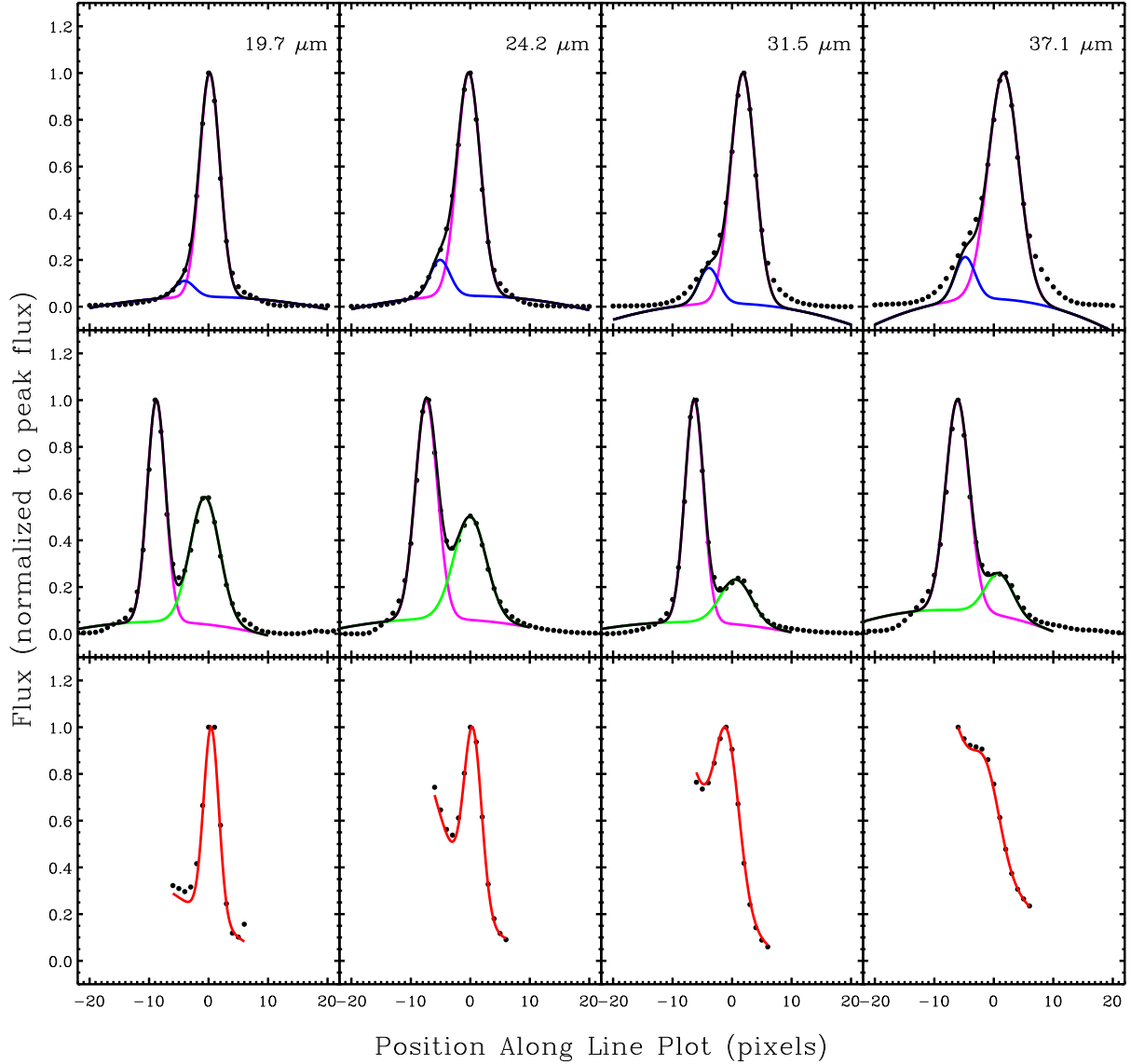}
\caption{1-dimensionally integrated and normalized line profiles (points) and best Gaussian fits for W3(OH) (magenta line), W3(H$_2$O) (blue line), 
the Northeastern H {\small II} region (green line), and the point source 2MASS J02270352+6152357 (red line). 
The 1-dimensional integrations were performed along the short directions of the boxes in Fig. \ref{fig:deconvolved_images_extract}. 
The black lines represent the sum of the fits. The extracted flux densities
for all sources are given in Table \ref{tab:fluxes}.
\label{fig:line_plots}}
\end{figure}

Fig.  \ref{fig:deconvolved_images_extract} shows boxed regions where the flux profiles for W3(OH) and W3(H$_2$O) were 
extracted (integrated vertically) and fit to Gaussian functions (Fig. \ref{fig:line_plots}).
Each of the profiles shows a secondary peak coincident with the location of W3(H$_2$O). The area under the
profile fit to W3(H$_2$O) yields its flux.  
The extracted flux densities for W3(H$_2$O) are also listed in Table \ref{tab:fluxes}. We discuss our detection of this source 
further in \S\ref{sec:w3h2o}. The flux densities for W3(OH) measured in this fashion agree to within $\sim 10\%$ of the
flux density measured from large aperture photometry, which is listed in Table \ref{tab:fluxes}.

Fig. \ref{fig:deconvolved_images_extract} also shows the regions selected for line profiles of the Northeastern H {\small II} region (integrated
horizontally) and 2MASS J02270352+6152357 (integrated vertically). 
The latter source is flagged in the 2MASS catalog as confused with neighboring objects; thus we do not consider its $2~\mu$m flux.
The line profiles for these 1-dimensional integrations are shown in Fig. \ref{fig:line_plots} and their
flux densities, determined by Gaussian line fits, are listed in Table \ref{tab:fluxes}.

\section{DISCUSSION}\label{Discussion}
\subsection{W3 (OH)}

\begin{figure}
\plotone{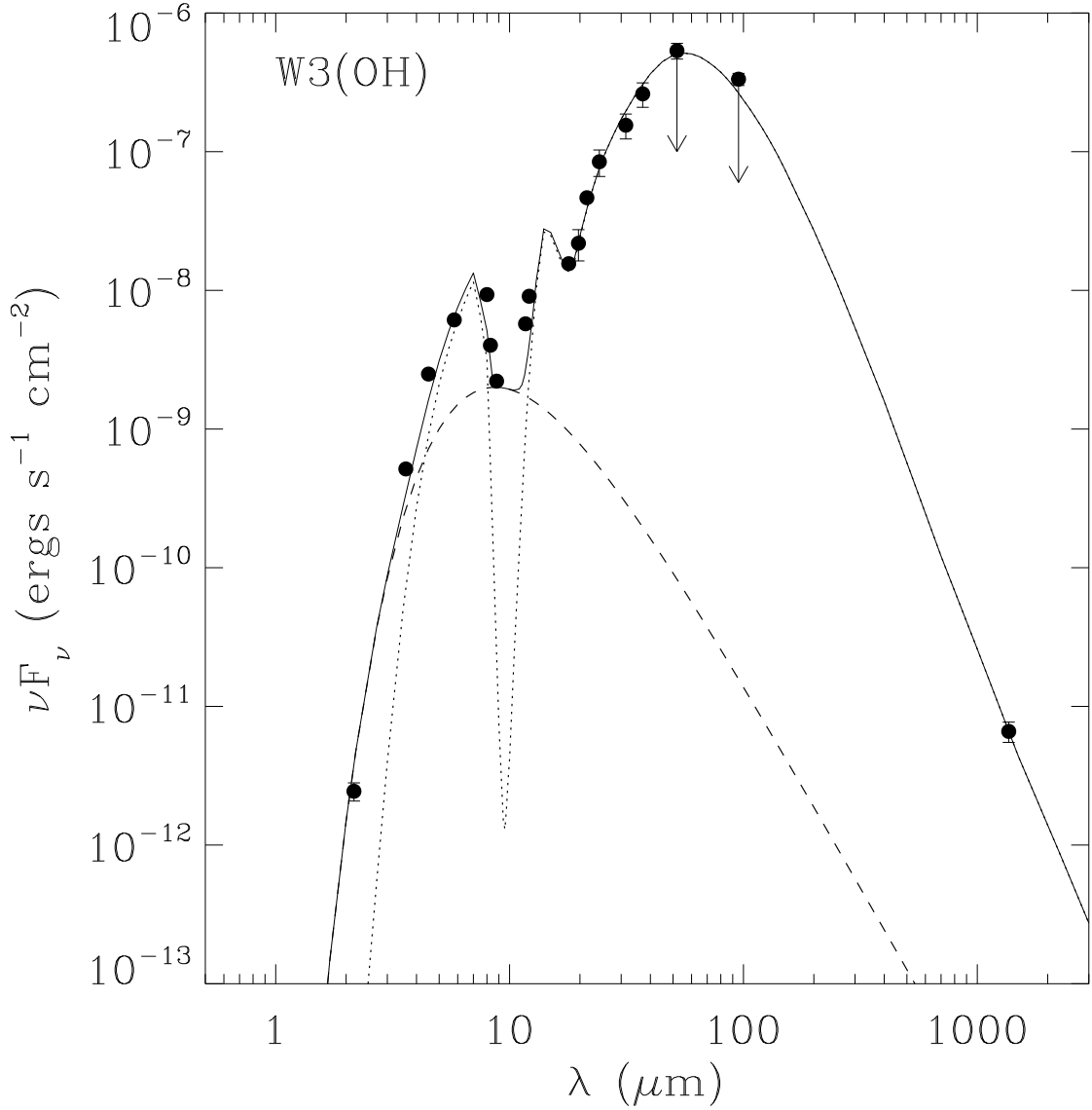}
\caption{SED of W3(OH) and model fits. {\it Filled circles:} Combined data points (see text for references). 
The IRAS fluxes were taken as upper limits, indicated as
downward arrows at $\sim 60$ and $100~\mu$m. 
{\it Dotted line}: DUSTY model SED.
{\it Dashed line}: Improvised single-temperature blackbody originating from 
interior, warm dust emission through a cavity opening.
{\it Solid line}: Total SED from DUSTY model and improvised single-temperature blackbody. 
\label{fig:w3ohsed}}
\end{figure}

In Fig. \ref{fig:w3ohsed} , we show the SED for W3 (OH) with additional data taken from the literature: 2MASS (2.2 $\mu$m), \citet{Stecklum}
(8.8, 12.7, and 17.8 $\mu$m), MSX ($21~\mu$m, Egan et al. \citeyear{egan99}), IRAS (60 and 100 $\mu$m), and \citet{Chini86} (1.3 mm). The IRAS flux densities were
taken as upper limits due to contamination from nearby sources.

We model the dust component as an optically thick, dusty shell around the H {\small II} region, irradiated at the inner boundary by the central star with surface
temperature $3.11 \times 10^4$ K.
The emergent SED of the model
was computed using the DUSTY
radiative transfer code\footnotemark \footnotetext[7]{Ivezic, Z., Nenkova, M., and Elitzur, M. 1996, University of Kentucky,
http://www.pa.uky.edu/\~moshe/dusty/}.
The density distribution of this model ($\rho \propto r^{-p}$; $p=1.5$) is based on a free-fall density profile \citep[e.g.][]{hartmann09}.
We chose the composition of the dust to be 53\% silicates and 47\% graphite grains \citep{draine84}.
We consider a grain size distribution $n(a) \propto a^{-q}$, whereby $q=3.5$ (Mathis, Rumpl, \& Nordsieck \citeyear{mathis77}; hereafter MRN), and with a minimum
grain size of 0.001 $\mu$m and a
maximum grain size of 0.25 $\mu$m (MRN; Sellgren \citeyear{sellgren84}).
We set the temperature of the inner edge of the shell at 400 K for this model. This is substantially
cooler than the dust sublimation temperature ($\sim 1500$ K), indicating the presence of a large cavity depleted of dust and consistent with previous work
\citep{Chini86,Stecklum}.
The total luminosity of this model is $\sim 7.1 \times 10^4~L_\odot$.
The SED of this model is shown in Fig. \ref{fig:w3ohsed} as the dotted line and the relevant model parameters are listed
in Table \ref{tab:w3ohmodel}.

Although they both use an optically thick, free-fall density shell in the IR, this model differs quantitatively from the model presented in Stecklum et al.
(\citeyear{Stecklum}). The central star is of later O-type and cooler surface temperature. The inner dust shell radius is nearly four
times smaller and the outer radius nearly twice as compact as in Stecklum et al. These parameters are necessary in order to explain the mid-IR emission in the range
$17-37~\mu$m. This work finds that the total luminosity is slightly lower than previous estimates.

Note this model cannot account for the excess emission at $2.2 - 5.8~\mu$m. One explanation for this emission
is clumpiness in the dust cloud, or a cavity opening, which allows short wavelength radiation from warm dust to leak through holes in the cloud.
We model radiation from the warmer interior of the shell as a single-temperature (425 K) blackbody component. The plausibility of this model
is evident in Fig. \ref{fig:w3ohsed}, whereby an excellent fit to the data is aChieved with a temperature consistent with
that of the interior of the dust shell. There may also be a scattered light component at these wavelengths, but
we do not model such a component since its geometry is relatively unconstrained.

An alternate explanation for the excess at $2.2-5.8~\mu$m is a relatively high abundance of 
very small grains compared with large grains.
In order to fit the data at $2.2-5.8~\mu$m, we would need to consider a model with a modified MRN grain size distribution ($q=4.7$).
However, in the ionized region, small grains can become super-heated and destroyed. The emission would come from
a relatively high abundance of small grains in the infalling shell; such grains would need to be primordial. 
Given the jets that are seen in molecular emission, it is most likely
that the excess short wavelength radiation originates in a cavity opening. The prediction one can make from this explanation is
the presence of continuum emission in the amporphous silicate absorption band at 9.7 $\mu$m, 
which decreases the depth of the absorption feature.
Mid-IR spectroscopy in the $8-13~\mu$m window could be used to test this prediction.

\subsection{W3(H$_2$O)}\label{sec:w3h2o}
In the radio continuum, the hot core W3(H$_2$O) consists of 3 clumps designated as A, B, and C \citep{Wyrowski99,Wilner}. 
An early claim to a detection of W3(H$_2$O) in the infrared came from  \citet{keto92} who presented
ground-based observations at $12.2~\mu$m
showing emission at the location of W3(H$_2$O) C, which lies approximately $4''$ East of W3(OH). No emission was
seen from sources A and B, which lie approximately $6''$ East of W3(OH). However, this observation was not
substantiated by \citet{Stecklum} who did not detect any emission from these sources at $8-12~\mu$m, despite their 
spatially resolving W3(OH).
In the FORCAST data, the separation of the peaks in the double-peaked line profile (Fig. \ref{fig:line_plots}) yields
the separation between W3(H$_2$O) and W3(OH), which we find to be $5.0''$ at 37.1 $\mu$m and $4.4''$ at 31.4 $\mu$m.
Components A and C form a massive protobinary system \citep{minh07}. Based on the gas chemistry
of these sources, \citet{minh07}  suggest that component A may be more deeply embedded and younger than
component C within $<10^4$ yr. This would be consistent with our results that show a wavelength dependence
for the $19.7 - 37.1~\mu$m emission. However, it should be noted that this dependence on wavelength may also result from
combined temperature and optical depth gradients.

\subsection{Protostars and Young Stars with Disks in the SOFIA Field}
\begin{figure}
\plotone{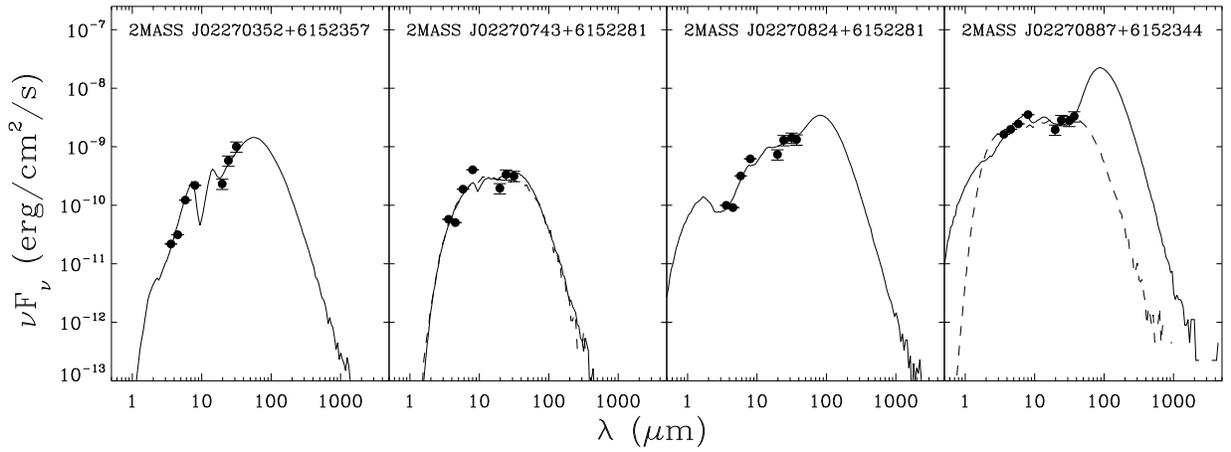}
\caption{SEDs of the four point sources in the SOFIA field (filled circles) and the best-fit protostellar SED model (solid line) from the online
SED fitting tool of \citet{Robitaille06,Robitaille07}. For  2MASS J02270743+6152281 and 2MASS J02270887+6152344, the best-fitting
alternative, disk-dominated model SEDs are also shown (dashed lines).
\label{fig:seds}}
\end{figure}

We construct SEDs for the four point sources in the FORCAST field and compare them with model SEDs of protostellar and young stellar
objects using the online SED fitting tool of \citet{Robitaille06,Robitaille07}. 
The SEDs and those of the best-fit models are shown in Fig. \ref{fig:seds}. We provide comments on each object in 
\S\ref{J02270352+6152357} -- \S\ref{J02270887+6152344}.

\subsubsection{2MASS J02270352+6152357}
\label{J02270352+6152357}
The best fitting model for J02270352+6152357 corresponds to a protostellar object with a mass of 3.96 M$_{\odot}$ and luminosity of 236 L$_{\odot}$. The model indicates that this object is young and cool, with an age of just under 5000 years and a temperature in the low 4000s of Kelvins. It is undergoing active envelope accretion from a massive envelope (11.7 M$_{\odot}$) onto a low-mass disk ($3.92 \times 10^{-3}$). The self-consistency of the top ten models with this best-fitting model suggest that this object is indeed a young, intermediate-luminosity protostar.

\subsubsection{2MASS J02270743+6152281}
\label{J02270743+6152281}
The SED fitting tool produced two families of best-fit models for J02270743+6152281, with the first family headed by the best-fitting SED and the second by the second-best fit. The families differ in the object's stage of 
envelope accretion and disk formation, with one family representing an object still actively undergoing envelope accretion 
and the other corresponding to a more highly-developed disk with a depleted envelope.

The best fitting (total $\chi^2 = 82.14$) model, representing the first family of models within the top ten best fits, 
indicates that this object is a protostellar object with a 
mass of 5.69 M$_{\odot}$ and a luminosity of 768 L$_{\odot}$, surrounded by a relatively substantial envelope (0.153 M$_{\odot}$) and a smaller disk ($3.36 \times 10^{-2}$ M$_{\odot}$). 
This model has a low but nonzero rate of envelope accretion, and suggests an age of $6.47 \times 10^{5}$ yr, with an internal temperature of 16,500 K. 
These parameters point to an intermediate-luminosity protostar nearing the end of its envelope accretion phase.

The second-best fit (total $\chi^2 = 82.72$), and correspondingly the second family of fits, represents an object with a fully formed disk and a completely depleted envelope no longer accreting material onto the disk. 
This model protostar has an intermediate mass somewhat higher than the alternative family, of 9.17 M$_{\odot}$. 
Its envelope is negligible ($10^{-5}$ M$_{\odot}$), and its disk has a mass of $2.1 \times 10^{-2}$ M$_{\odot}$. 
It suggests an age of just over $10^6$ yr and a temperature of 24,500 K. Given the protostellar mass and these parameters, this family of models likely corresponds to a young star with a circumstellar disk.

The two families of models differ primarily in the presence or absence of an amorphous silicate absorption feature at 9.7 $\mu$m, which is
typically seen in an envelope-dominated SED. Highly sensitive mid-infrared spectroscopy could be used to resolve the degeneracy between 
these two families of models.

\subsubsection{2MASS J02270824+6152281}
\label{J02270824+6152281}
Model fitting for J02270824+6152281 presents a convergent set of parameters. The best fit model corresponds 
to a 6.11 M$_{\odot}$ protostar with a luminosity of 432 L$_{\odot}$. 
This model suggests the object is a young protostar (approximately 8000 years) with a temperature of around 4000 K. 
The object is embedded in a large 35.1 M$_{\odot}$ envelope with a disk of mass $1.48 \times 10^{-2}$ M$_{\odot}$, 
indicating that this object is likely experiencing ongoing disk formation. 
The relatively high envelope accretion rate displayed by this model ($1.39 \times 10^{-3}$ M$_{\odot}$ yr$^{-1}$) 
supports this conclusion. 
The models predict substantial far-IR emission arising from the envelope, meaning followup observations in the far-IR range could confirm our assessment.

\subsubsection{2MASS J02270887+6152344}
\label{J02270887+6152344}
For J02270887+6152344, the data again result in two possible families of models, representing either an intermediate- to 
high-mass protostellar object with a large envelope or an older intermediate-mass young star with a clearly defined disk and minimal envelope. The top three fits all corresponded to the former of these models, indicating its greater likelihood for accuracy.

The best-fitting (total $\chi^2 = 15.03$) model for this source represents a 10.5 M$_{\odot}$ protostar surrounded by a natal dust envelope containing 240 M$_{\odot}$ of material and a negligible disk. The model object has a luminosity of 3000 L$_{\odot}$. An age of approximately 2000 yr and a temperature in the 4000s of Kelvins 
indicate that the object is protostellar in nature.

The second group of well-fitting models is represented by the fourth-best fitting (total $\chi^2 = 34.34$) model. This model corresponds to a young star of mass 9.8 M$_{\odot}$, with a depleted envelope (mass approximately $2 \times 10^{-5}$ M$_{\odot}$, indicating that the envelope has nearly entirely accreted onto the disk). This model possesses an age of over $10^6$ years and an internal 
temperature of 25,400 K, indicating a more advanced stage in development than the first family of models presented.

These families of models diverge at far-infrared wavelengths. 
With only present data, we therefore cannot rule out either the higher-mass, younger protostellar object or the intermediate-mass, 
more evolved young star. Further observations of this object at far-IR and/or submillimeter wavelengths are required to resolve the degeneracy in the model parameters.

\section{CONCLUSIONS}
We present SOFIA/FORCAST and Spitzer/IRAC observations of the UCH {\small II} region W3(OH) in the wavelength range $3.6 - 37.1~\mu$m. 
These data, combined with other published data, have been used to constrain the optical depth, grain size distribution, and 
temperature gradient in the dusty shell
surrounding the H {\small II} region. The total luminosity of W3(OH) is $7.1 \times 10^4~L_\odot$, indicating that the central
star is an O9 star with surface temperature $\sim 31,000$ K. A clumpy dust distribution or cavity opening revealing warm interior 
grains is necessary to explain excess emission at $2.2 - 4.5~\mu$m. 

We detect the hot core W3(H$_2$O) at 31.4 and $37.1~\mu$m, and constrain its flux density at $19.7 - 37.1~\mu$m 
using deconvolved FORCAST images. 

In addition, SEDs have been constructed for four young stellar or protostellar objects
which lie in the SOFIA/FORCAST field. The model SED fitting tool of \citet{Robitaille06} was used to determine the
nature of these objects. 
2MASS J02270352+6152357 is an intermediate-luminosity protostar undergoing envelope accretion;
2MASS J02270824+6152281 
is most likely a very young intermediate-mass protostar with a large natal envelope;
2MASS J02270887+6152344 
is a high-luminosity object which is either a protostar with ongoing envelope accretion onto a young disk
or a young star with a circumstellar disk and a depleted envelope; 
 and 
2MASS J02270743+6152281 could be an intermediate-luminosity protostar or potentially a young star with a 
developed disk and an almost entirely depleted envelope.  Further observations in the mid-IR, far-IR and/or
submillimeter range(s) are required to definitively characterize 2MASS J02270887+6152344  and 
2MASS J02270743+6152281.

\acknowledgments
We thank R. Grashius, S. Adams, H. Jakob, A. Reinacher, and U. Lampeter for their SOFIA telescope engineering and operations support. We also thank
the SOFIA flight crews and mission operations team (A. Meyer, N. McKown, C. Kaminski) for their SOFIA flight planning and flight support.
We are grateful to an anonymous referee for his or her comments which have improved this manuscript.
This work is based on observations made with the NASA/DLR Stratospheric Observatory for Infrared Astronomy (SOFIA). SOFIA science mission operations are conducted
jointly by the Universities Space Research Association, Inc. (USRA), under NASA contract NAS2-97001, and the Deutsches SOFIA Institut (DSI) under DLR contract 50
OK 0901. Financial support for FORCAST was provided to Cornell by NASA through award 8500-98-014 issued by USRA.
This work is based in part on observations made with the Spitzer Space
Telescope, which is operated by the Jet Propulsion Laboratory, California
Institute of Technology under a contract with NASA.
This publication makes use of data products from the 
Two Micron All Sky Survey, which is a joint project of the University of Massachusetts 
and the Infrared Processing and Analysis Center, funded by the National Aeronautics and 
Space Administration and the National Science Foundation.
This research has made use of the NASA/ IPAC Infrared Science ArChive, which is operated by the Jet Propulsion Laboratory, California Institute of Technology,
under contract with the National Aeronautics and Space Administration.
This research has made use of NASA's Astrophysics Data System Abstract Service. \\

{\facility {\it Facilities}: 2MASS, {\it Spitzer Space Telescope}, SOFIA, IRAS}

\newpage

\newpage

\newpage

\newpage

\newpage

\newpage

\newpage

\newpage

\newpage
\begin{deluxetable}{lccccccc}
\setlength{\tabcolsep}{0.02in}
\rotate
\tablewidth{0pt}
\tabletypesize{\small}
\tablecaption{\label{tab:fluxes}Positions and flux densities in Janskies for W3(OH), W3(H$_2$O), and four mid-IR sources in the SOFIA field.}
\tablehead{\colhead{} & \colhead{J02270352+6152357} & \colhead{W3(OH)} & \colhead{Northeastern H {\tiny II}} & \colhead{W3(H$_2$O)} & \colhead{J02270743+6152281} & \colhead{J02270824+6152281} & \colhead{J02270887+6152344} }
\startdata
RA (2000)    & 02 27 03.52 & 02 27 03.83  & & & 02:27:07.43 & 02:27:08.24 & 02:27:08.87 \\
DEC (2000)  & +61 52 35.7 & +61 52 24.8 & & & +61:52:28.1 & +61:52:28.1 & +61:52:34.4   \\
%$F_{1.66}$  & & & \nodata & $2.46\times10^{-2}\pm1.47\times10^{-3}$ & $2.68\times10^{-3}\pm1.4\times10^{-4}$ & $7.93\times10^{-3}\pm2.58\times10^{-4}$  \\
%$F_{2.16}$  & & & $1.76\times10^{-3} \pm 2.36\times10^{-4}$ & $0.155\pm0.0$ & \nodata & \nodata  \\
$F_{2.16}$  & \nodata & $1.76\times10^{-3} \pm 2.36\times10^{-4}$ & \nodata & \nodata & \nodata & \nodata & \nodata \\
$F_{3.6}$   & $0.0262\pm0.00013$ & $0.619\pm0.00027$ & \nodata & \nodata & $0.0691\pm0.026$ & $0.120\pm0.00018$ & $1.96\pm0.00027$ \\
$F_{4.5}$  & $0.0470\pm0.000088$ & $3.74\pm0.00018$ & \nodata & \nodata & $0.0759\pm0.026$ & $0.137\pm0.00012$ & $2.97\pm0.00018$ \\
$F_{5.8}$  & $0.237\pm0.00054$ & $11.8\pm0.0011$ & \nodata & \nodata & $0.364\pm0.15$ & $0.613\pm0.00072$ & $4.76\pm0.0011$ \\
$F_{8.0}$		& $0.584\pm0.00030$ & $24.9\pm0.00060$ & \nodata & \nodata & $1.08\pm0.41$ & $1.66\pm0.00040$ & $9.43\pm0.00060$ \\
$F_{19.7}$  & $1.53\pm0.31$ & $144\pm37$ & $19.7\pm3.9$ & $9.41\pm1.9$ & $1.28\pm0.26$ & $4.86\pm0.97$ & $12.8\pm2.6$ \\
$F_{24.2}$  & $4.68\pm0.94$ & $683\pm148$ & $79.1\pm16$ & $76.7\pm15$ & $2.70\pm0.54$ & $10.5\pm2.1$ & $23.1\pm4.6$ \\
$F_{31.4}$  & $10.6\pm2.1$ & $1626\pm332$ & $170\pm34$ & $147\pm29$ & $3.32\pm0.66$ & $15.0\pm3.0$ & $29.0\pm5.8$ \\
$F_{37.1}$  & $13.0\pm2.6$ & $3232\pm646$ & $223\pm45$ & $225\pm45$ & \nodata & $16.4\pm3.3$ & $40.9\pm8.2$ \\
\enddata
\end{deluxetable}

\newpage
\begin{deluxetable}{lc}
\tabletypesize{\small}
\tablecaption{\label{tab:w3ohmodel}DUSTY parameters for best-fit model of the W3(OH) dust component. The parameters are
stellar temperature $T_*$; inner shell radius $R_{in}$; outer shell radius $R_{out}$, inner shell boundary temperature $T_{in}$,
outer shell boundary temperature $T_{out}$, grain composition, density distribution parameter $p$, grain size distribution parameter $q$,
and optical depth $\tau_{37}$ at $37~\mu$m.}
\tablehead{\colhead{Parameter} & \colhead{Value}}
\startdata
$T_{*}$       & 31145 K \\
$R_{in}$      & 576 AU   \\
$R_{out}$    & 29942 AU \\
$T_{in}$      & 400 K \\
$T_{out}$    & 26 K \\
Composition fraction, silicates & 53\% \\
Composition fraction, graphite & 47\% \\
$p$             &  1.5 \\
$q$             & 3.5 \\
$\tau_{37}$ & 2.8 \\
\enddata
\end{deluxetable}

\end{document}